\documentclass{ws-procs9x6}
\def\g2{{$(g-2)$} }
\begin{document}

\title{Results and Future Prospects for Muon $(g-2)$
\footnote{\uppercase{T}his work is supported in part by
the \uppercase{U.S.} \uppercase{N}ational \uppercase{S}cience 
\uppercase{F}oundation and the \uppercase{U.S.} \uppercase{D}epartment of 
\uppercase{E}nergy.}}

\author{B.~Lee ROBERTS\footnote{\uppercase{O}n behalf of
the \uppercase{M}uon $(g-2)$ 
\uppercase{C}ollaboration.$^{11,12,13}$}}

\address{Department of Physics \\
Boston University \\ 
590 Commonwealth Avenue\\
Boston, MA 02215 USA\\ 
E-mail: roberts@bu.edu}

\maketitle

\abstracts{Spin physics had its beginnings in the famous experiments of 
Stern and Gerlach, which eventually resulted in the postulation of
spin by  Goudsmit and Uhlenbeck.  The Stern-Gerlach experiment told
us that the $g$-value of the electron was 2, but we now know
that because of radiative corrections, the $g$-value of the leptons
is slightly greater than 2, the lowest-order contribution being
$\alpha/\pi$, where $\alpha$ is the fine-structure constant. 
Measurements of the magnetic dipole moments of the electron and
muon have played a major role in our understanding of QED and
of the standard model.  In this talk
I discuss the progress on measurements and theory of the 
magnetic dipole moment of the  muon.
}

\section{Theory of the Lepton Anomalies}

Over the past 83 years, the study of dipole moments of 
elementary particles has provided a wealth of information on 
subatomic physics, and more recently has provided topics of interest to
this conference.   The pioneering work of Stern\cite{stern}
led to the discovery of spin, and
showed that $g_e\simeq 2$.  This set the stage
for the precision measurements by Foley and Kusch,\cite{kusch}
which showed $g$ was not exactly 2, but rather slightly larger,
which was explained by Schwinger\cite{schwinger} and played
an important role in the development of QED.  Subsequently Stern\cite{sternp}
showed that  $g_p\simeq 5.5$, and Alvarez and Bloch\cite{nmdm} found
that the neutron had a magnetic moment, which 
eventually helped lead to the quark models of the baryons.
In the 1980s, measuring hyperon magnetic moments to test
quark models became an
industry that was well covered in earlier installments of
these spin conferences, and in which I had the pleasure of participating.

A charged particle with spin $\vec s$ has a magnetic moment
\begin{equation}
 \vec \mu_s = g_s ( {e \over 2m} ) \vec s;
\qquad   a \equiv { (g_s -2) \over 2};\qquad   \mu = (1 + a){e \hbar \over 2m};
\end{equation}
where $g_s$ is the gyromagnetic ratio, $a$ is the anomaly,
and the latter expression is what one finds in the Particle
Data Tables.\cite{pdg}

For point particles, the anomaly arises from radiative corrections.
The QED contribution to {$a$} (or $g$) is an expansion 
in $ \left({\alpha \over  \pi}\right)$,
$a =\sum_{n=1} C_n \ \left( {\alpha \over  \pi}\right)^n$,
with one diagram for the Schwinger (second-order) contribution
(where $C_1 = 0.5$),
five for the fourth order, 40 for the sixth order, 891 for the eighth
order.
The QED contributions to electron and muon \g2 have now been calculated
through eighth order, $(\alpha/\pi)^4$, and the 
tenth-order contribution has been estimated.\cite{kinqed}

The electron anomaly is measured to a relative precision
of about 4 parts in a 
billion (ppb),\cite{eg2} which is better than the precision on 
the fine-structure constant $\alpha$, and
Kinoshita has used the measured 
electron anomaly to give the best determination
of $\alpha$.\cite{kinalpha} The electron anomaly will be further improved
over the next few years.\cite{gab}
 
The muon anomaly has been measured to 0.5 parts per 
million (ppm).\cite{brown2,bennett1,bennett2}  The relative contributions of 
heavier particles to $a$ scales as $(m_e/m_{\mu})^2$, so the muon
has an increased sensitivity to higher mass scale radiative corrections
of about 40,000 over the electron.
At a precision of $\sim 0.5$ ppm, the muon anomaly
is sensitive to $\geq 100$ GeV scale physics.

The standard model value of $a_{\mu}$ has 
measurable contributions from three types of radiative
processes: QED loops containing leptons ($e,\mu,\tau$) and 
photons;\cite{kinqed}
hadronic loops containing hadrons in vacuum polarization 
loops;\cite{davmar,dehz2,HMNT,HICHEP,hlbl}
and weak loops involving the $W$ and $Z$ weak gauge bosons
(the standard model Higgs contribution is negligible),\cite{davmar}
$a_{\mu}{\rm ( SM)}
= a_{\mu}({\rm QED}) + a_{\mu}({\rm Had}) +
a_{\mu}({\rm Weak})$.
A significant 
difference between the experimental value and the standard model
prediction would signify the presence of new physics.
A few examples of such potential contributions are
lepton substructure, anomalous $W-\gamma$
couplings, and supersymmetry.\cite{davmar}

\begin{figure}[h!]
\begin{center}
  \includegraphics[width=.55\textwidth]{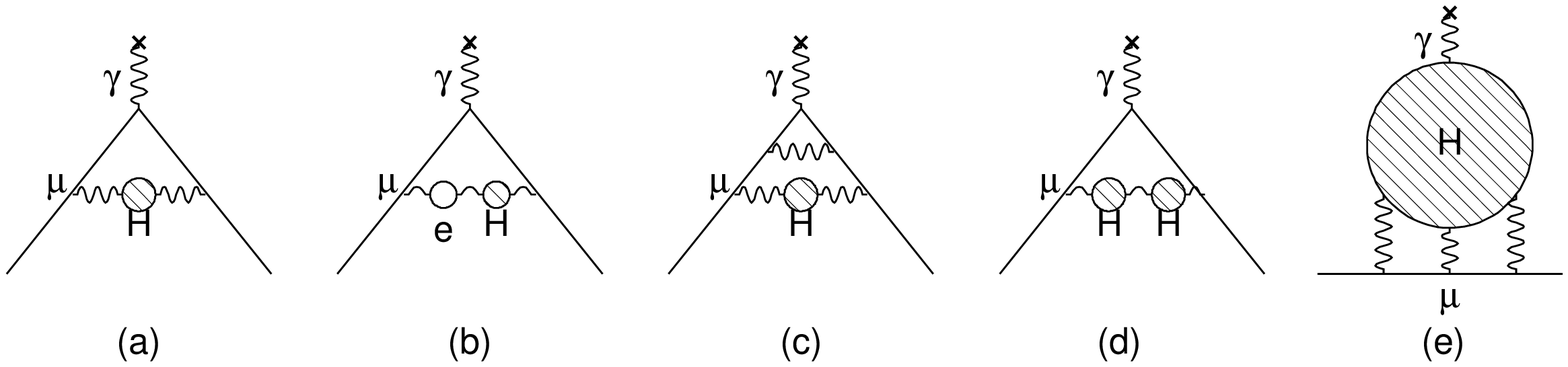}
\end{center}
  \caption{The hadronic contribution to the muon anomaly, where the
dominant contribution comes from (a).  The hadronic light-by-light
contribution is shown in (e).}
\label{fg:had}
\end{figure}

The CERN experiment\cite{cern3}
observed the contribution of hadronic vacuum polarization 
shown in Fig. \ref{fg:had}(a)  at the
8 standard deviation level.  Unfortunately, the hadronic contribution
cannot be calculated directly from QCD, since the energy scale
is very low ($m_{\mu} c^2$), although Blum\cite{blum} has performed 
a proof of principle calculation on the lattice.  
Fortunately dispersion theory 
gives a relationship between the vacuum polarization loop
and the cross section for $e^+ e^- \rightarrow {\rm hadrons}$,
\begin{equation}
a_{\mu}({\rm Had;1})=({\alpha m_{\mu}\over 3\pi})^2
\int^{\infty} _{4m_{\pi}^2} {ds \over s^2}K(s)R(s);
\ \ R\equiv{ {\sigma_{\rm tot}(e^+e^-\to{\rm hadrons})} \over
\sigma_{\rm tot}(e^+e^-\to\mu^+\mu^-)}
\end{equation}
and experimental
data are used as input.
The factor $s^{-2}$ in the dispersion relation, means that values
of  $R(s)$ at low energies (the $\rho$ resonance) dominate the determination of
$a_{\mu}({\rm Had;1})$. In principle, this information could 
be obtained from hadronic $\tau^-$ decays such as
$\tau^- \rightarrow \pi^- \pi^0 \nu_{\tau} $, which can be 
related to $e^+e^-$ annihilation through the CVC hypothesis and
isospin conservation.\cite{dehz2} 
However, inconsistencies between
information obtained from $e^+e^-$ annihilation and hadronic
tau decays, plus an independent 
confirmation of the CMD2  high-precision $e^+e^-$ cross-section
measurements by the KLOE collaboration,\cite{KLOE} 
have prompted Davier, H\"ocker, et al.,  to state that 
until these inconsistencies can be understood
only the $e^+e^-$ data should be used to 
determine $a_{\mu}({\rm Had;1})$.\cite{HICHEP}

The hadronic light-by-light contribution (see Fig. \ref{fg:had}(e)) 
has been the topic of much
theoretical investigation.\cite{hlbl}  Unlike the lowest-order
contribution, it can only be calculated from a model, and this contribution
is likely to provide the ultimate limit to the precision of the
standard-model
value of $a_{\mu}$.

One of the very useful roles the measurements of $a_{\mu}$ have played in the 
past is placing serious restrictions on physics beyond the standard model.
With the development of supersymmetric theories as a favored scheme of
physics beyond the standard model, interest in the experimental and
theoretical value of $a_{\mu}$ has grown substantially.  Contributions 
to $a_{\mu}$ from
SUSY or other new dynamics at the several hundred GeV scale
 could be at a measurable level in a broad range of models.
Furthermore, there is a complementarity between the SUSY contributions 
to the magnetic (MDM) and electric dipole (EDM) moments
 and the transition moment for the lepton-flavor
violating (LFV) process $\mu^- \rightarrow e^-$ in the field of a nucleus.  
The MDM and EDM are related to the real and
imaginary parts of the diagonal element of the slepton 
mixing matrix, and the transition moment is related to the
off-diagonal one. See Klaus Jungmann's talk from this conference for
a discussion of electric dipole moments.

\section{Measurement of the muon anomaly}
The method used in the third
CERN experiment and the BNL experiment are
very similar, save the use of direct muon 
injection\cite{kick} into the storage ring,\cite{mag,inf}
which was developed by the E821 collaboration.  These
experiments are based on the
fact that for  $a_{\mu} > 0$ the spin 
precesses faster than
the momentum vector when a muon travels transversely to a 
magnetic field.  The spin precession frequency $\omega_S$
consists of the Larmor and Thomas spin-precession terms. The
spin frequency $\omega_S$, the momentum   
precession (cyclotron) frequency $\omega_C$,  are given by 
\begin{equation}
 \omega_S = {geB \over 2 m c} + (1-\gamma) {e B \over \gamma mc};\ 
 \omega_C = {e B \over mc \gamma}; \ \ 
\omega_a = \omega_S - \omega_C = \left({g-2 \over 2}\right) {eB \over mc}.
\label{eq:omegas}
\end{equation}
The difference frequency $\omega_a$
is the frequency with which the spin
precesses relative to the momentum, and is  proportional to
the anomaly, rather than to $g$.
A precision measurement of $a_{\mu}$ requires precision measurements
of the muon spin precession frequency $\omega_a$,  and the magnetic field,
which is expressed as the free-proton precession frequency
$\omega_p$ in the storage ring magnetic field.

The muon frequency can be measured as accurately as the counting
statistics and detector apparatus permit.  
The design goal for the NMR magnetometer and calibration system
was a field accuracy of 0.1 ppm.  The $B$ which enters in 
Eq. \ref{eq:omegas} is the average field seen by the ensemble of muons
in the storage ring.  In E821 we reached a precision of 0.17 ppm in the
magnetic field measurement.

An electric quadrupole field\cite{quads} is used for vertical focusing, 
taking advantage of the 
``magic''~$\gamma=29.3$ at which an electric field does not contribute to
the spin motion relative to the momentum. With both an electric
and a magnetic field, the spin difference frequency is given by
\begin{equation}
\vec \omega_a = - {e \over mc}
\left[ a_{\mu} \vec B -
\left( a_{\mu}- {1 \over \gamma^2 - 1}\right) \vec \beta \times \vec E
\right],
\label{eq:tbmt}
\end{equation}
which reduces to Eq. \ref{eq:omegas} in the absence of an electric field.
For muons with $\gamma = 29.3$ in an electric field alone,
the spin would follow the momentum vector.

\begin{figure}[h!]
\begin{center}
  \includegraphics[width=.65\textwidth]{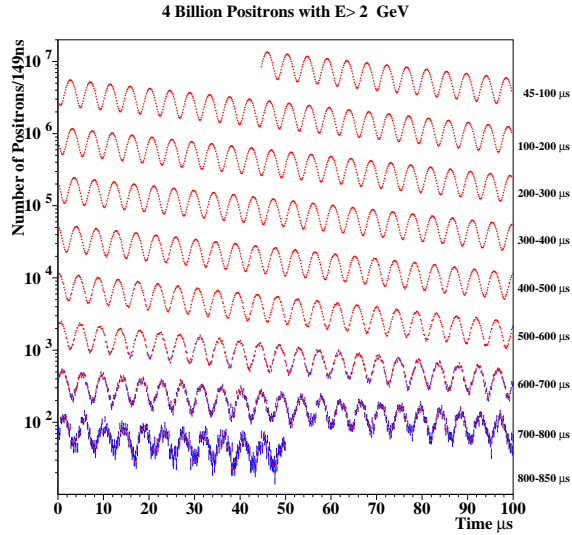}
\end{center}
  \caption{The time spectrum of positrons with energy greater than
2.0 GeV from the year 2000 run.  The endpoint energy is 3.1 GeV.
The time interval for each of the diagonal ``wiggles'' is given
on the right.}
\label{fg:wig00}
\end{figure}

The experimental signal is the $e^{\pm}$ from $\mu^{\pm}$ decay, which 
were detected by lead-scintillating
fiber calorimeters.\cite{det}  The time and energy of each event was
 stored for analysis offline. 
Muon decay is a three-body decay, so the 3.1 GeV muons produce a continuum
of positrons (electrons) from the end-point energy down.  Since the highest
energy  $e^{\pm}$ are correlated with the muon spin, if one counts high-energy 
 $e^{\pm}$ as a function of time, one gets an exponential from muon decay
modulated by the $(g-2)$ precession. The expected form for the positron time
spectrum is $f(t) =  {N_0} e^{- \lambda t } 
[ 1 + {A} \cos ({\omega_a} t + {\phi})] $, however in analyzing the
data it is  necessary
to take a number of small effects into account in order to obtain
a satisfactory $\chi^2$ for the fit.\cite{bennett1,bennett2}
The data from our 2000 running period are shown in Fig. \ref{fg:wig00}

\begin{figure}[h!]
\begin{center}
  \includegraphics[width=.45\textwidth]{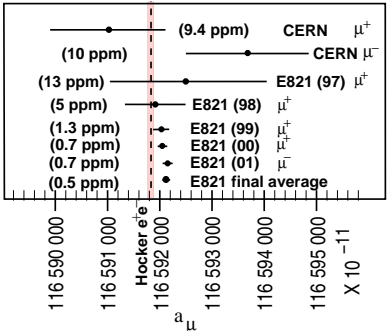}
\end{center}
  \caption{Measurements of $a_{\mu}$ compared with the
theory value described
in the text.
 }
\label{fg:amu}
\end{figure}

The experimental results from E821 are shown in Fig. \ref{fg:amu}, with
the average 
\begin{equation}
a_\mu(\rm{E821}) = 11\,659\,208(6) \times 10^{-10}
\qquad (\pm 0.5\ {\rm ppm})
\end{equation}
which determines the ``world average''.  The theory 
value\cite{kinqed,davmar,HICHEP}
$a_\mu(\rm{SM}) = 11\,659\,182.8(7.3) \times 10^{-10}$, 
$(\pm 0.7\ {\rm ppm})$ is determined using the strong interaction
contribution
from H\"ocker et al.,\cite{HICHEP} which updates 
their earlier analysis\cite{dehz2} with the KLOE data.\cite{KLOE}
The value of
Hagiwara et al.,\cite{HMNT} gives an equivalent answer.  
The hadronic light-by-light contribution of
$(12.0 \pm 3.5)\times 10^{-10}$
is taken from Davier and Marciano\cite{davmar}.
When the experimental value is compared to the standard model value
using either of these two analyses\cite{HMNT,HICHEP}
for the lowest-order hadronic
contribution, one finds
$\Delta a_{\mu}({\rm E821-SM}) = 
( 25.2 \ {\mathrm to }\ 26.0 \pm 9.4  )\times 10^{-10}$, 
(2.7 standard deviations).

To show the sensitivity of our measurement of
$a_{\mu}$ to the presence of virtual electroweak gauge bosons, we
subtract off the electroweak contribution of 
$ 15.4 (0.1)(0.2) \times 10^{-10}$
from the standard model value, compare with experiment and obtain
$\Delta a_{\mu} = (40.6 \pm 9.4)  \times 10^{-10}$,
a 4.3 standard deviation discrepancy.  This difference shows conclusively
that E821 was sensitive to physics at the 100 GeV scale.  At present,
it is inconclusive whether 
we see evidence for contributions from physics beyond the standard-model
gauge bosons.

With each data set, the systematic error was reduced, as can be seen
from Table \ref{tb:ervrst}, and the experiment was statistics limited
when running was ended.  Given the tantalizing discrepancy 
between our result and the latest standard-model
value, and the fact that the hadronic error could be
reduced by about a factor of two over the next few years,\cite{davmar} 
we submitted a
new proposal to Brookhaven to further improve the experimental measurement.
The goal of this new experiment is $\pm 0.2$ ppm total error, with the
goal of controlling the total systematic errors on the magnetic field
and on the muon frequency measurement to 0.1 ppm each.

\begin{table}[h!]
\centering
\tbl{Systematic and statistical errors from the three major E821 data \break
runs. }
{\small
\begin{tabular}{||l|c|c|c|c||} \hline
Data  &{$\omega_p$} ($B$-Field) & {$\omega_a$ } &{Total }        & { Total }\\
 Run  &  {Systematic}  & {Systematic }   & { Systematic   } & {Statistical } \\
      & {Error (ppm)}  &{  Error (ppm)}  & { Error (ppm) }  & { Error (ppm)} \\
\hline
1999 & 0.4  &  0.3 &{0.5} &{ 1.3} \\
2000 & 0.24 &  0.31 &{0.39} &{ 0.62} \\
2001 &0.17 & 0.21 & {  0.27} &{0.66} \\
\hline
E969 & 0.1 & 0.1 & 0.14 & 0.14 \\
Goal &     &     &      &      \\
\hline
\end{tabular}}
\label{tb:ervrst}
\end{table}

Our proposal\cite{E969} was given enthusiastic scientific approval in 
September 2004 by the Laboratory, and has been given the new
number, E969. Negotiations are underway between the Laboratory and
the funding agencies to secure funding. 

The upgraded experiment will use a backwards muon beam to reduce background
in the electron calorimeters. A new inflector magnet with open ends will
be employed. The beamline improvements will increase the stored flux in the
ring by $\sim 5$, and the detectors, electronics and data acquisition system
will be replaced with components which can handle the increased rates
with reduced systematic errors.  

\begin{figure}[h!]
\begin{center}
  \includegraphics[width=\textwidth]{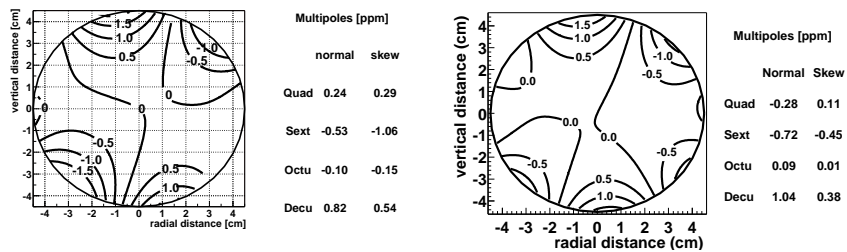}
\end{center}
  \caption{The magnetic field averaged over azimuth in the storage ring
for the 2000 ($\mu^+$) (left figure) and 2001 ($\mu^-$) (right figure)
running periods.  Contours represent
0.5 ppm changes.  The multipole content is in ppm relative to the dipole
field.
}
\label{fg:field}
\end{figure}

In E821, the magnetic field was uniform to about one ppm, as can be seen from
Figure \ref{fg:field}.  To improve our knowledge of the field from 0.17 ppm
to 0.1 ppm, we will further shim the storage ring and improve on
the calibration, monitoring and measurement of the magnetic field.

A letter of intent (LOI) for an even more precise \g2 experiment was also
submitted to J-PARC.\cite{g2jparc}  In that LOI we proposed to reach a
precision below 0.1 ppm.  Since it is not clear how well the hadronic
contribution can be calculated, and whether the new Brookhaven experiment
E969 will go ahead, we will evaluate whether to press forward with this
experiment at a later time.  Our LOI at J-PARC\cite{g2jparc} 
was predicated on
pushing as far as possible at Brookhaven before moving to Japan.

\section{Summary and Conclusions}

Muon \g2 has played an important role in constraining the standard model
for many years.   With the sub-ppm accuracy now available for the 
muon anomaly,\cite{brown2,bennett1,bennett2} 
there may be indications that new physics is beginning to
appear in loop processes.\cite{mar}  
An enormous amount of work continues worldwide to improve on our knowledge
of the hadronic contribution, and we can look forward to a factor of about
two improvement over the next few years.  We have proposed to improve on 
the precision of the measurement by a factor of two and a half.  These
two improvements will provide a much more sensitive confrontation with the
standard model in the next few years.

{ \section* {Acknowledgments}}
I wish to thank my colleagues on the muon \g2 experiment, as well as  
M. Davier, J. Ellis, T. Kinoshita, E. de Rafael,  
W. Marciano and T. Teubner  for helpful discussions. Special thanks to
Dave Hertzog, Jim Miller and Yannis 
Semertzidis for helpful comments on this manuscript.


\begin{thebibliography}{9}
\bibitem{gab} G. Gabrielse and collaborators have constructed a new trap, and
expect to improve on the electron anomaly by a factor of 10 to 100 over
the measurements of Ref. 8.
\end{thebibliography}
\end{document}